# Terahertz Radiation Power Characterization and Optimization of Stack of Intrinsic Josephson Junctions

Alireza Kokabi, Hamed Kamrani, and Mehdi Fardmanesh, *Senior Member, IEEE*

*Abstract*— Terahertz radiation of the stack of intrinsic Josephson junctions in the mesa structure of the layered high-$T_c$ superconductors is analyzed and presented in this work. The dependency of the radiated power to the geometrical parameters, cavity-waveguide boundaries, and magnetic and electric bias has been investigated. This has been done by numerical calculation of the previously proposed coupled sine-Gordon equations, which characterize the electromagnetic dynamics of the stack of the intrinsic Josephson junctions. Using the obtained numerical results from these coupled equations, the effect of the design parameters such as dimensions of the mesa structure and the magnitude of the applied magnetic field and the dc current on the enhancement of the radiated power is studied. Thus, the radiated power is optimized with respect to these considered parameters. By variation of the number of layers, we also investigate the effect of the number of intrinsic Josephson junctions on the total radiated power. The results from this part are also compared with the previous analytical models.

*Index Terms*—Terahertz radiation, Stack of intrinsic Josephson junctions, Flux-flow radiation

## I. Introduction

DESPITE of being an extremely attractive band, the "terahertz gap" which is assumed from 0.5 to 10THz is hardly applicable because of the lack of practical sources in this range [1]. However, the stack of intrinsic Josephson junctions in the mesa structure of layered high-$T_c$ superconductors is a promising candidate that could fill this gap [2]. In such structures, synchronization of high-frequency oscillations in all Josephson junctions can produce terahertz radiation [3].

A very promising practical way of applying this synchronization in finite-size mesas is the excitation of internal cavity resonance that can be followed by a large number of junctions through a chain reaction [4,5]. In this mechanism, which is known as Fiske mode, the frequency of such an in-phase resonance is controlled by the width of the junctions. For efficient coupling, this lateral size should be small while the number of the junctions should be very large.

Theoretical possibility of terahertz emission from intrinsic Josephson junctions has been investigated more than a decade [6-17]. First, Josephson-vortex flow was proposed as the operation principle of experimentally observed emission in the presence of the applied magnetic field [18]. However, the emission of electromagnetic wave without applied magnetic field is also observed in the GHz region [19].
It is well known that there exist various dynamic states for the system in the radiation mode. The power of the emitted electromagnetic radiation in the coherent emission or super-radiation regime is known to be proportional to $N^2$, $N$ being the number of junctions.

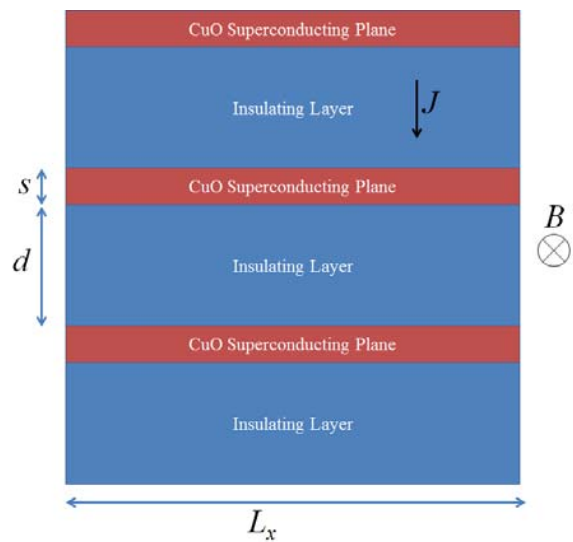

Fig. 1. Schematic diagram of the emitter of terahertz electromagnetic wave

Depending on the length of the junctions, there exist two kinds of oscillatory regimes. In the case of system of short junctions satisfying $L_x < 3\lambda_c$, we deal with monochromatic standing-wave oscillation at the Josephson frequency. On the other hand, in the system of long junctions with $L_x > 3\lambda_c$, vortex-antivortex excitations lead to solitonic modes and

Manuscript received November 8, 2011.
A. Kokabi is with the Department of Electrical Engineering, Sharif University of Technology, Tehran, Iran, and P.O. Box: 11365-11155 (corresponding author to provide phone: +98-21-66165989; fax: +98-21-66165990; e-mail: kokabi@ee.sharif.ir).
H. Kamrani is with the Department of Electrical Engineering, Sharif University of Technology, Tehran, Iran, and P.O. Box: 11365-11155.
M. Fardmanesh is with the Department of Electrical Engineering, Sharif University of Technology, Tehran, Iran, and P.O. Box: 11365-11155.

nonlinearity effects whose fundamental harmonic is equal to the half Josephson frequency.

By now, theoretical calculations [3,18] as well as numerical simulations [10,11,20-27] are presented which investigate possible radiation from the stack of intrinsic Josephson junction. Current-voltage characteristic is another interested topic in this area. For instance, charging effects of CuO layers on the current-voltage characteristics of the stacked junctions are considered previously [28,29]. A method is also proposed for the synchronization of all junctions for the whole region of I-V characteristics in addition to the resonance regions [30]. A wide range of different effects has been investigated in these works. However, many issues have not yet been revealed concerning the mechanisms and characterizations of the terahertz emission. For instance, the optimum radiation power with respect to the involving parameters such as magnitude of the applied magnetic field, sample width, and electrical bias is not fully concerned.

The experimental works show that the observed radiated power is very low in the order of 100W/cm$^2$ by Kadowaki [31] *et al.* and 20W/cm$^2$ by Bae *et al.*[19]. Thus, the optimization of this power would be of high interest in order to achieve the maximum power efficiency. In this case, the dependency of the terahertz intensity with respect to some of the size parameters are investigated previously [32]. Considering these parameters such as bias current and cavity-waveguide boundary size along with consideration of other parameters such as number of layers in the mesa structure and the magnitude of the applied magnetic field, here we investigate the optimization of the emitted power. The case of non-magnetically biased is also considered in this works. We use the Koshelev model [18] to compute the radiation modulated voltage response of the stack of intrinsic Josephson junctions in the single crystal of the high-$T_c$ superconductor BSCCO. The proposed model is expressed as a system of $N$ coupled differential sine-Gordon equations. The problem is solved with the dynamic boundary conditions of the radiated electromagnetic wave.

## II. THEORETICAL MODEL

We start from the definition of gauge-invariant phase difference as [4-6]

$$\psi_{l+1,l}(\mathbf{r},t) = \phi_{l+1}(\mathbf{r},t) - \phi_l(\mathbf{r},t) - \frac{2\pi}{\Phi_0}\int_{z_l}^{z_{l+1}} A_z(\mathbf{r},t)dz, \quad (1)$$

with the phase of order parameter in the $l$-th superconducting layer being $\phi_l$, where $A_z$ stands for the vector potential, the quantum flux is expressed by $\Phi_0$, and $z_l$ is the height of the $l$-th superconducting layer. In this paper, the subscript "$l+1,l$" denotes quantities in the insulating layer between the $l$-th and ($l+1$)th superconducting planes. We assumed that the thicknesses of the superconducting and insulating layers are $s$ and $d$ respectively.

Electromagnetic dynamics of gauge-invariant phase difference in the presence of applied magnetic field is described as a set of coupled sine-Gordon equations that correspond to the total pair and quasi-particle tunneling. According to the Josephson relation, the total current density that passes through each junction along the z-axis can be obtained from

$$J_{l,l+1} = J_c \sin\varphi_{l+1,l} + \sigma_c E_{l+1,l} + \frac{\varepsilon}{4\pi}\frac{\partial}{\partial t}E_{l+1,l}, \quad (2)$$

where $J_c$ is the Josephson critical current density, $\sigma_c$ is the normal conductivity along the c-axis, and $E_{l+1,l}$ is the z-axis component of the electric filed between ($l$+1)th and $l$-th layers. The c-axis conductivity, $\sigma_c$, is temperature dependent, and the damping term can be neglected in small values of $\sigma_c$. In the above equation, the first, second, and third terms represent the pair tunneling, quasi-particle tunneling, and the charge accumulation currents respectively. Using the London theory as well as the Maxwell equations, the properties of the system can be expressed as [27]

$$\frac{\partial^2\varphi_{l+1,l}}{\partial\chi^2} = \left(1 - \zeta\Delta_n^2\right)\left(\frac{\partial\xi_{l+1,l}}{\partial\tau} + \beta\xi_{l+1,l} + \sin\varphi_{l+1,l} - j\right)$$

$$\frac{\partial\varphi_{l+1,l}}{\partial\tau} = \left(1 - \alpha\Delta_n^2\right)\xi_{l+1,l}, \quad (3)$$

in which the finite difference operator is defined as $\Delta_n^2 f_n = f_{n+1}+f_{n-1}-2f_n$, and the parameters $\alpha=\varepsilon_c\mu^2/sd$ and $\zeta=\lambda_{ab}/sd$ are capacitive and inductive couplings between different junctions and $\beta=4\pi\sigma_c\lambda_c/\varepsilon_c^{1/2}$ is the damping term. In the definition of these parameters, $\lambda_c$ and $\lambda_{ab}$ are London penetration depth along the c-axis and CuO planes respectively, $\varepsilon_c$ and $\mu$ are the relative permittivity and susceptibility inside the cavity respectively, and $\sigma_c$ is the quasiparticle conductivity across the c-axis. We have replaced the quantities $\psi_{l+1,l}$, $E_{l+1,l}$, $x$, and $t$ by their equivalents $\varphi_{l+1,l}$, $\xi_{l+1,l}$, $\chi$, and $\tau$ in an scaled unit which is defined by following transformations,

$$\chi = \frac{x}{\lambda_c}, \tau = \omega_p t, j = \frac{J}{J_c}, \xi_{l+1,l} = \frac{\sigma_c}{\beta J_c}E_{l+1,l}, \quad (4)$$

where $J_c$ and $\omega_p$ are the critical current and plasma frequency of the intrinsic junctions that can be replaced by $c\Phi_0/8\pi^2\lambda_c^2 d$ and $c/\varepsilon_c^{1/2}\lambda_c$ respectively. In such a reduced unit, the scaled magnetic field, $b_{l+1,l}$, is also obtained from

$$\frac{\partial\varphi_{l+1,l}}{\partial\chi} = \left(1 - \zeta\Delta_n^2\right)b_{l+1,l}, \quad (5)$$

while it is related to the magnetic field through $b_{l+1,l}=2\pi\lambda_c dB_{l+1,l}/\Phi_0$. The dynamical boundary condition [8] between the oscillatory parts of scaled boundary fields is applied on the edges with the assumption of infinite number of junctions, following Lin *et al.* [24] leading to

$$\frac{\partial\varphi_{l+1,l}}{\partial\chi} = \langle b_{l+1,l}\rangle_\tau + \tilde{b}_{l+1,l}$$

$$\frac{\partial\varphi_{l+1,l}}{\partial\tau} = j/\beta + \tilde{\xi}_{l+1,l} \quad (6)$$

$$\tilde{b}_{l+1,l} = \pm z\sqrt{\varepsilon_c/\varepsilon_d}\tilde{\xi}_{l+1,l},$$

where $z$ models the impedance mismatch between the cavity and the dielectric, $\varepsilon_d$, is the relative permittivity of the dielectric, the symbol $\langle\ldots\rangle_\tau$ stands for time averaging, and the symbol ~ over field quantities represents their oscillatory parts.

The final equation approximates the boundary condition proposed in ref. [18] in the special case of infinite number of the junctions where the nonuniformity of oscillations can be neglected. Such a condition is not satisfied in the characterization of the radiation with respect to the number of junctions. In this case, the original boundary conditions should be applied.

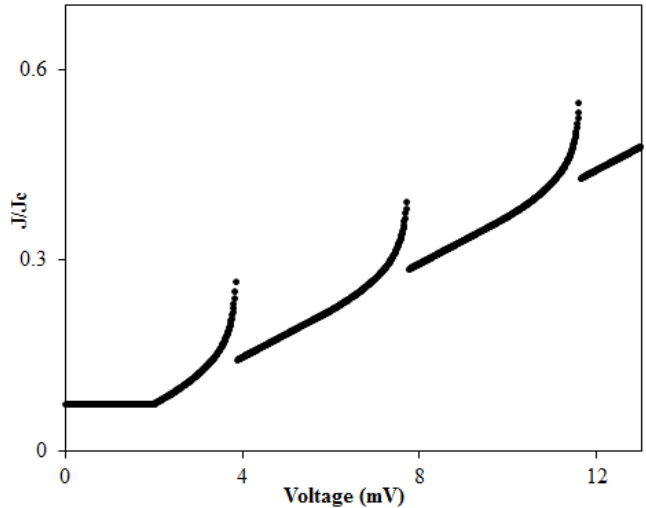

Fig. 2. Current-voltage characteristics for $L_x$=80μm

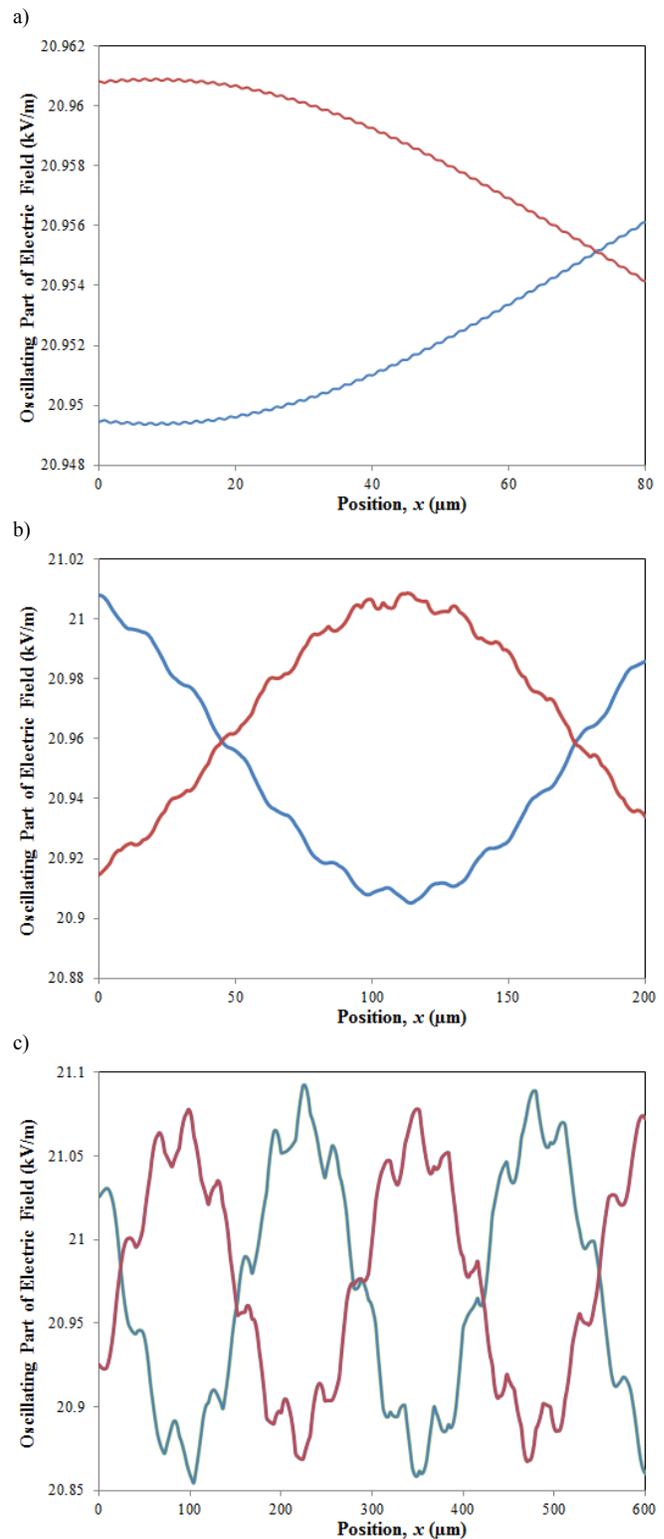

Fig. 3. Position variation of the electric field inside the cavity for the case of a) $L_x$=80μm, b) $L_x$=200μm and c) $L_x$=600μm all at $B$=1T and $J/J_c$=0.1

The initial configuration for the electric field, $E_{l+1,l}$, is set to be its final dc value while the zero initial condition is applied for the gauge-invariant phase difference. These initial values result in the faster convergence as well as preventing from being trapped by the local minima. The voltages over all junctions are also assumed to be equal in correspondence with previous numerical works [20, 23]. It should also be noted that the magnetic modulation of the critical current is automatically included in the simulations [33-35].

### III. RESULTS AND ANALYSES

Equation (3) is solved for $N = 20$ to $60$ superconducting layers. The values of $\lambda_c=200\mu m$, $\omega_p=4.7\times10^{11}$rad/s, $s=3$Å, $d=12$Å, $\alpha=0.1$, $\beta=0.02$, $\zeta=5\times10^5$, and $\varepsilon_c=10$ are chosen for the parameters, which are very similar to the ones of BSCCO superconductor. However, our simulations show negligible dependency of results to capacitive and inductive coupling parameters in almost wide range of variations. Thus, similar analysis is applicable to the other types of layered superconductors. In the numerical simulations, the relative tolerance of the results between successive time steps is set to be lower than $10^{-7}$. The mesh size of $0.001$ in the reduced unit of length is selected. The radiation intensity is calculated for the obtained results by the method similar to the previous works [19]:

$$S_i = \frac{\varepsilon_d c}{2}\left(\tilde{E}_{l+1,l}\right)^2 \tag{7}$$

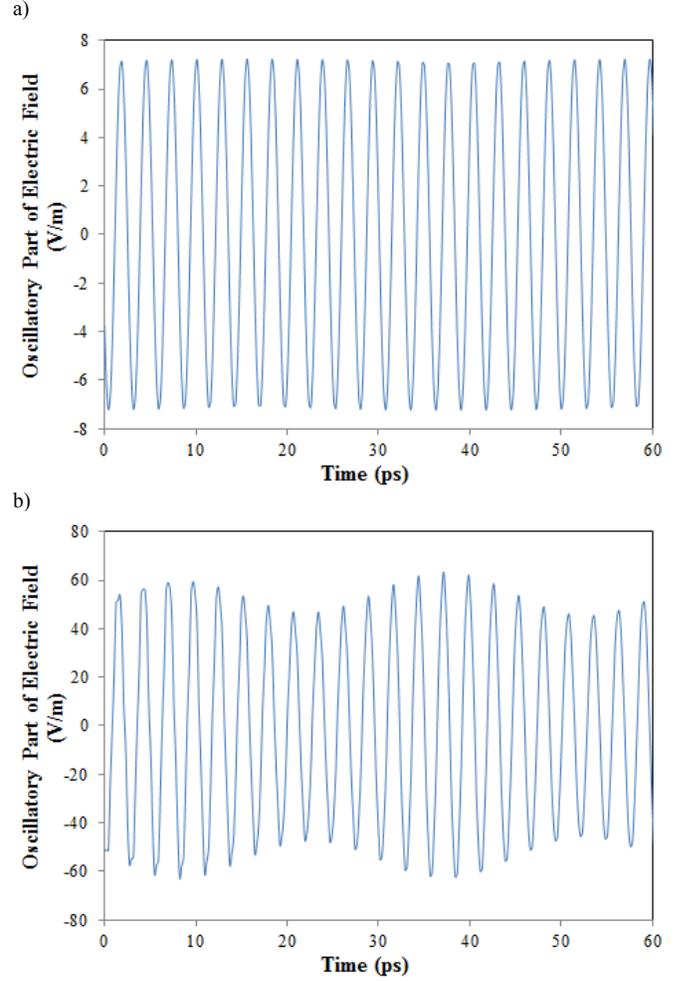

Fig. 4. Time evolution of the radiated electric field inside the dielectric for a) $L_x=20\mu m$ and b) $L_x=600\mu m$ all at $B=1$T and $J/J_c=0.1$

#### A. Sample results

Fig. 2 presents the current-voltage characteristics obtained for $L_x=80\mu m$ at the absence of external magnetic field. In Fig. 3, the distribution of the oscillating part of the electric field over the position in different times is plotted. Fig. 3.a shows this distribution for the cavity size of $L_x=80\mu m$. We expect standing waves inside the cavity for this size. Fig. 3.a confirms this prediction and the standing waves are clearly observed. It is also expected that by increasing cavity size, this stationary condition moves toward solitonic regime. Fig. 3.b shows the electric field distribution for the case of $L_x=200\mu m$ which is almost an intermediate situation between solitonic and standing wave regimes. Finally, one can observe that more than one harmonic is excited in the part c of this figure obtained for the sample width of $600\mu m$, which is soliton-like behavior.

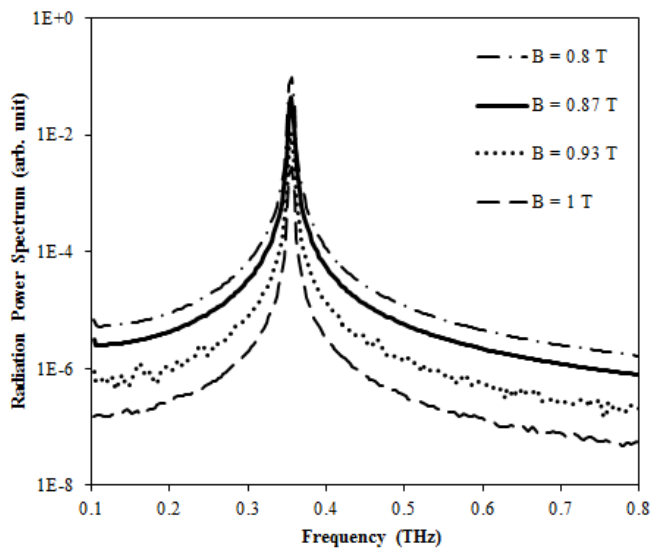

Fig. 5. Radiation power spectrum for different values of applied dc magnetic fields ($J/J_c$=0.2, $L_x$=40μm)

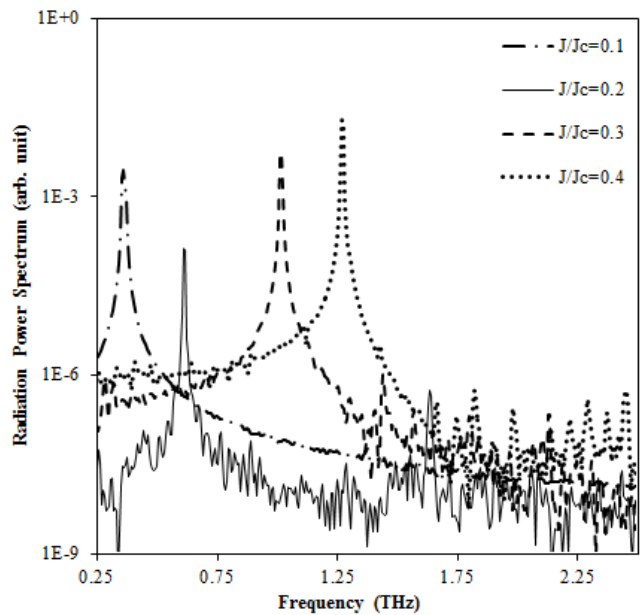

Fig. 6. Radiation power spectrum for different values of *c*-axis bias currents ($B_a$=1T, $L_x$=40μm)

The results of investigation for the time evolution of the electric field at the boundaries of the cavity are shown in Fig. 4. The first part of this figure shows the time evolution for the case of $L_x$=20μm while Fig 4.b is obtained for a sample width of $L_x$=600μm. The former has monochromatic behavior, while the latter is modulated by low frequency harmonics.

### B. Power Spectrum Characterization

In this section, the variation and optimization of the radiated power at different design parameters is discussed. We assumed a square form sample with the magnetic field in direction of *a*-axis while the effects of finite size in the *b*-axis are neglected.

In Fig. 5, the power spectrums of the emitted electromagnetic wave for different values of applied magnetic field are presented. Since Fourier analysis is performed on the finite number of data points, some fluctuations appeared in the spectrums. As illustrated in this figure, in the considered range, the magnetic field does not affect on the center frequency of the radiation. However, decreasing the magnetic field slightly broadens this spectrum and increases the adjacent harmonics.

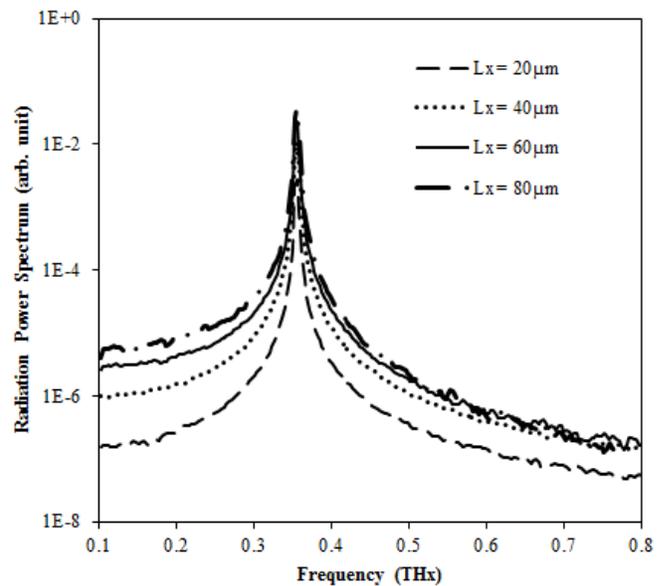

Fig. 7. Radiation power spectrum for different values of cavity widths ($B_a$=1T, $J/J_c$=0.2)

TABLE I
SIMULTANEOUS ANALYSIS OF RADIATION POWER VERSUS MAGNETIC FIELD AND CURRENT ($L_x$=20μm)

|  | $B_a$=0 | $B_a$=0.8T | $B_a$=0.87T | $B_a$=0.93T | $B_a$=1T |
|---|---|---|---|---|---|
| $J/J_c$=0.1 | 650nW | 702nW | 335nW | 210nW | 83.7nW |
| $J/J_c$=0.2 | 738nW | 643nW | 198nW | 194nW | 192nW |
| $J/J_c$=0.3 | 844nW | 528nW | 79.1nW | 42.6nW | 317nW |
| $J/J_c$=0.4 | 421nW | 390nW | 96.7nW | 137nW | 428nW |

In Fig. 6, the power spectrums of emitted wave for different applied bias currents are depicted. The figure shows that increasing the bias current shifts the center frequency to higher values. This might be associated to the increase of Lorentz force applied to the Josephson vortices from the *c*-axis current that make them move faster. The faster move Josephson vortices, the rapider change the magnetic field and hence the frequency increases. The figure also shows that the frequency of the peak intensity in the power radiation spectrum almost linearly depends on the applied dc current. In addition, excluding the case of $J/J_c$=0.2, the peak power shows ascending behaviors in the both domains of the electrical bias and the frequency. However, when $J/J_c$=0.2 is applied, its associated peak is slightly lower than other ones while an extra peak is also observed at about 1.5THz in the spectrum. It seems that in this special case, two harmonics are excited and the second one absorbed a fraction of the radiation power. It might be associated to the interference between traveling adjacent Josephson vortices at this special current. The behavior is similar to the one of an assumptive case in which semi-vortices move between the two neighboring Josephson vortices and swing the magnetic field with a duplicated frequency at a lower power.

The dependency of the power spectrum to the cavity width is presented in Fig. 7. As it is concluded from the figure, though that widening the cavity would not change the center frequency, it strongly enhances the adjacent harmonics especially in the low frequency side of the central harmonic. As we previously mentioned, low-dimension cavities results in the generation of stationary waves and monochromatic behavior, while wide cavities produce solitonic waves with large nonlinearity and side harmonics. Widening of the spectrum with increasing cavity size in Fig. 7 further confirms the above analyses.

TABLE II
SIMULTANEOUS ANALYSIS OF RADIATION POWER VERSUS CAVITY WIDTH AND CURRENT ($B$=0.87T)

|  | $L_x$=20μm | $L_x$=40μm | $L_x$=60μm | $L_x$=80μm |
|---|---|---|---|---|
| $J/J_c$=0.1 | 83.4nW | 271nW | 428nW | 30.4nW |
| $J/J_c$=0.2 | 320nW | 440nW | 66.2nW | 75.1nW |
| $J/J_c$=0.3 | 554nW | 64.0nW | 474nW | 139nW |
| $J/J_c$=0.4 | 485nW | 276nW | 212nW | 538nW |

*C. Radiation Power Optimization*

Three major parameters, which are cavity width, bias current, and applied dc magnetic field might affect on the radiated power. In order to find the optimized point for the output power, one should obtain the dependency of power to these parameters. In the following, we perform the optimization process with respect to each pair of these parameters while the full 3D optimization that needs a large amount of processing is under further investigations.

In tables I, II and III, the variations of the emitted power with respect to each pair of the mentioned parameters are shown. In the considered grids, which contain values in the order of experimental samples, one could observe a semi-oscillatory behavior of the electromagnetic power as the cavity size, bias current and the dc applied magnetic field change. Maximum power among these grids is obtained with $L_x$=20μm, $B$=0T, and $J/J_c$=0.3. On the other hand, simulations show that minimum power is generated by a sample of $L_x$=20μm width, exposed to the dc magnetic field of $B$=0.93T, and biased by the dc current of $J/J_c$=0.1.

Due to the heating effect, temperature control, and efficiency considerations, it would be of great interest to study the fraction of the radiated power to the power fed into the samples, $Q=P_{rad}/P_{in}$, for these simulation results. While $P_{in}$ is proportional to $J^2 L_x^2$, by looking into the tables I to III, one can find that the maximum and minimum radiation power efficiencies are obtained by the parameters ($J/J_c$=0.3, $B$=0T, $L_x$=20μm) and ($J/J_c$=0.1, $B$=0.87T, $L_x$=80μm) to be 3% and 0.006% respectively. Thus, applying the magnetic field of $B$=0.8T and the $c$-axis current of $J/J_c$=0.1 to a sample of width $L_x$=20μm leads to the maximum radiation power along with the maximum electrical power efficiency, simultaneously.

Similar to these numerical results for the radiation power can be found in the previously published experimental data. As reported by Ozyuzer *et. al.*, 20μW of pumping power is observed for 500 junctions [4]. Assuming that the number of junctions quadratically controls the power, this result is comparable with our data obtained for 60 junctions, regardless of temperature effects. More recently, 20μW at the radiation peak is observed for a rectangular mesa sample of 1μm thickness. Such a thickness contains about 700 junctions which is about 40% more than the previous one [36]. Scaling this power by quadratic relation corresponds to 150nW for 60 junctions. Since the sample is rectangular (300μm×80μm) and the exact details of the electrical bias is not mentioned in the paper, it cannot directly be compared with the simulation data. However, the value is close to the result obtained for a square mesa of 80μm×80μm.

TABLE III
SIMULTANEOUS ANALYSIS OF RADIATION POWER VERSUS CAVITY WIDTH AND MAGNETIC FIELD ($J/J_c$=0.1)

|  | $B_a$=0 | $B_a$=0.8T | $B_a$=0.87T | $B_a$=0.93T | $B_a$=1T |
|---|---|---|---|---|---|
| $L_x$=20μm | 650nW | 702nW | 335nW | 21.0nW | 83.7nW |
| $L_x$=40μm | 332nW | 21.8nW | 626nW | 416nW | 271nW |
| $L_x$=60μm | 784nW | 694nW | 220nW | 184nW | 438nW |
| $L_x$=80μm | 201nW | 439nW | 30.4nW | 275nW | 434nW |

In Fig. 8, the dependency of the total radiated power intensity to the number of the layers is investigated. The figure shows that increasing the number of the mesa layers would increase the total power. As mentioned earlier, two regimes of operation are proposed for the radiation of the stack of Josephson junctions [18]. Synchronization of the junctions leads to super-radiation regime, which enhance the radiation power by the factor of $N^2$, $N$ being the number of synchronized junctions inside the space of the radiation wavelength. However, assuming independent junctions, the total power would be enhanced by $N$. In Fig. 8, these two extreme cases have been depicted. The simulation results are between these two extremes and closer to the synchronized regime. Since the calculations are performed for the finite number of junctions, the numerical results do not exactly coincide with the predictions for the synchronized systems. It is remarkable that in this figure, the quadratic and linear curves are fitted to the calculation results at the junction number of 20 in order to allow better comparison between these three plots in the considered range.

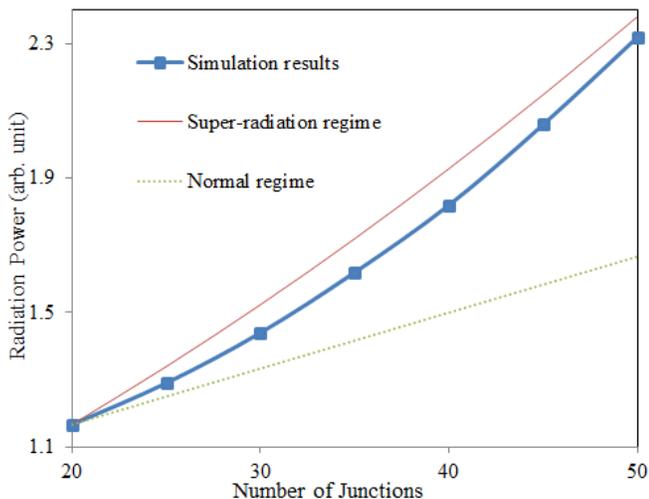

Fig. 8. Variation of the radiation power by change in the number of junctions ($B_a$=0.87T, $L_x$=20μm)

## IV. CONCLUSION

The terahertz radiation from the stack of intrinsic Josephson junctions is characterized in this paper. The power spectrum for different design parameters is obtained and discussed. Maximum radiation power and power efficiency are also investigated with respect to these parameters. The results show a semi-oscillatory behavior of the radiation power with respect to the magnetic field, bias current, and cavity size. In our analyses, the maximum radiation power and efficiency occur simultaneously in a special value of these parameters. Moreover, at some current biases more than one harmonic is excited.

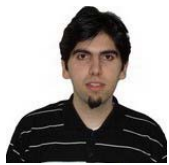

**Alireza Kokabi** was born in Iran in 1982. He received the B.S. and M.S. degrees in Electrical Engineering from Sharif University of Technology in 2006. He is currently working towards the Ph.D. degree in Electrical Engineering from Sharif University of Technology, Tehran, Iran. He is a member of the Superconductive Electronics Research Laboratory (SERL) of Electrical Engineering Department of Sharif University of Technology. His research interests include superconductivity, cryogenic system design, superconductor bolometers, SQUID devices, SQUID based systems, and solid state and photonic devices.

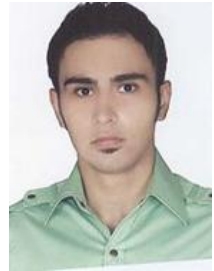

**Hamed Kamrani** was born in Iran in 1987. He received his B.Sc. degree in Electrical Engineering from Shahid Beheshti University in 2010. He is currently M.S Student in Sharif University of Technology. He is a member of the Superconductive Electronics Research Laboratory (SERL) of Electrical Engineering Department of Sharif University of Technology. His research interests include superconductivity, proton precession magnetometer, THz sources, and solid state and photonic devices.

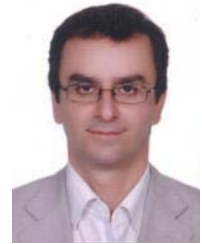

**Mehdi Fardmanesh** was born in Tehran, Iran in 1961. He received the B.S. degree from Tehran Polytechnic University and the M.S. and Ph.D. degrees from Drexel University, Pennsylvania, all in electrical engineering, in 1987, 1991, and 1993, respectively. In 1989, he joined Drexel University, and until 1993, he conducted research in development of the thin- and thick- film high temperature superconducting materials, devices, and development of ultralow noise cryogenic characterization systems, where he was awarded a research fellowship by the Ben Franklin Superconductivity Center in 1989. From 1994 to 1996, he was Principal Manager for R&D and the Director of a private sector research electrophysics laboratory, while also teaching at the Departments of Electrical Engineering and Physics of Sharif University of Technology in Tehran. In 1996, he joined the Electrical and Electronics Engineering Department at Bilkent University in Ankara, teaching in the areas of solid-state, and electronics, while also supervising his established Superconductivity Research Laboratory. In 1998 and